**Seismogenic nodes as a viable alternative to seismogenic zones and observed seismicity for the definition of seismic hazard at regional scale.**

Paolo Rugarli[a], Franco Vaccari[b,*] and Giuliano Panza[c,d,e,f]

a - CASTALIA S.r.l., Via Pinturicchio, 24, 20133, Milano, Italy.
b - Univ. di Trieste, Dip. di Matematica e Geoscienze, Trieste, Italy
c - Accademia Nazionale dei Lincei, Roma, Italy
d - Earthquake Administration, Beijing, China
e - International Seismic Safety Organization (ISSO), Arsita, Italy.
f - Beijing University of Civil Engineering and Architecture (BUCEA)
* - Corresponding author, vaccari@units.it

**Abstract**

It is shown that considering a fixed increment of a given magnitude at a fault is equivalent to factoring the mechanical moment at the fault as done in structural engineering with the applied loads, by the most currently used structural engineering standards (e.g. Eurocodes). A special safety factor $\gamma_{EM}$ is introduced and related to the partial factor $\gamma_q$ acting on the mechanical moment representing the fault.

A comparison is then made between the hazard maps obtained with the Neo Deterministic Seismic Hazard Assessment (NDSHA) technique, using two different approaches for the definition of the seismic sources considered for the computation of the synthetic seismograms.

The first one is based on the magnitude of the events, listed in the parametric earthquake catalogue compiled for the study area, and located within the active seismogenic zones. This is the standard approach that has been used in most of the NDSHA computations performed up to now. It is an adequate approach for countries, like Italy, where the catalogue completeness criterion, based on the validity of the loglinear Gutenberg-Richter relation (GR) as a law, is reasonably fulfilled for events of magnitude M≥5, since year 1000.

When this condition is not satisfied, i.e. when the catalogue completeness is barely adequate, a second approach can be adopted for the definition of the earthquake sources. It uses the seismogenic nodes identified in the region by means of pattern recognition techniques applied to morphostructural seismic zonation (MSZ), and increases the reference magnitude by a constant variation tuned thanks to the safety factor $\gamma_{EM}$.

The two approaches have been compared for Italy using $\gamma_{EM}$=2.0: in most of the territory they produce comparable hazard maps. As the two sets are totally independent and the Italian catalogue is very long, this implies a validation of the seismogenic nodes method and a tuning of the safety factor $\gamma_{EM}$ at about 2. Notable exception is seen in the Central Alps, where nodes tend to overestimate the "observed" hazard. Probably the current seismic activity as measured today, over a time interval of about 1000 years, may be well not representative of longer periods of time, on account of the peculiar local mechanical and rheological properties of the lithosphere in the area. In southeastern Sicily, the nodes underestimation is only apparent and can be negligible within experimental errors.

**Keywords**

Seismogenic nodes; Seismogenic zones; Maximum Credible Earthquake; Neo Deterministic Seismic Hazard Assessment; Eurocodes.

## 1 - Introduction

In the original formulation of NDSHA (Panza et al. 2001; 2012), physics-based computer computation was combined with a comprehensive geologic and geophysical overview of the regional tectonic setting and earthquake history to solve, in a *first approximation*, the fundamental problems posed by an adequate description of the physical process of earthquake occurrence (which in the real earth is a *tensor* phenomenon). It examines the largest scenario event physically possible, usually termed Maximum Credible Earthquake (MCE), whose cellular magnitude $M_{design}$ at a given site can be tentatively, until proven otherwise, set equal to the *maximum observed* or estimated magnitude $M_{max}$, plus some multiple of its accepted global standard deviation $\sigma_M$. In areas where information on faults and other input data are sparse, the historical data *together with* morphostructural analysis are relied upon to estimate this maximum magnitude (e.g. Parvez et al., 2017; Hassan et al., 2017; Rugarli et al., 2018).

According to *Chebyshev's theorem* for a very wide class of probability distributions, no more than a certain fraction of values can be more than a certain distance away from the *mean*. Specifically, no more than $1/k^2$ of a distribution's values can be more than $k$ standard deviations away from the *mean* (or equivalently, at least $1-1/k^2$ of the distribution's values are within $k$ standard deviations of the mean). If $k=2$, then at least 75% of the values fall within $2\sigma_M$ and if $k=3$ at least 89% of the values fall within an interval of $3\sigma_M$ centered on the *mean*.

The factor $k$ can be considered a *tunable* safety factor that may be applied coherently with the safety factors used in structural engineering, e.g. naming it $\gamma_{EM}$ (EM=Earthquake Magnitude).

So $M_{design}=M_{max}+\gamma_{EM}\sigma_M$, where it is currently assumed $\sigma_M=0.2-0.3$, and it is proposed to use $\gamma_{EM}=1.5-2.5$. Since the design value $M_{design}$ is determined by adding a further tunable increment to the *maximum* estimated value $M_{max}$ (not the *mean*), it must be considered an *envelope* ─ evaluated at the best of our present-day knowledge.

## 2 - Safety factors

In the mechanical systems currently used by engineers to evaluate the safety of the structures, the semi-probabilistic or partial safety factors paradigm has emerged as a reference in the last 40 years. In the following, reference will be made to the so called Eurocode 0 (CEN EN-1990:2002), which is a standard accepted worldwide, because the paradigm is described in detail. This standard is the basis of all the other Eurocodes, referring to actions (EN 1991), concrete structures (EN 1992), steel structures (EN 1993), seismic design (EN 1998) and so on.

Both the applied actions and the resistances are usually evaluated statistically, and *characteristic* values are computed. Characteristic values are values that, assuming some distribution of probability, have a given low probability to be exceeded (the actions), or to be unreached (the material resistances), in a given reference period or after some production process.

Among the existing actions, Eurocode 0 enlists also the so-called *accidental* actions, among which earthquakes are to be considered. An accidental action is defined in Eurocode 0 (CEN EN-1990:2002, par.1.5) as

> *an action, usually of short duration but of significant magnitude, that is unlikey to occur on a given structure during the design working life*
>
> *NOTE 2 Impact, snow, wind and seismic actions may be variable or accidental actions, depending on available information on statistical distribution.*

Eurocode 0 is aware that it is not always possible to have reliable statistics referring to actions. For these cases it admits that *the characteristic value of variable actions* is expressed as (Eurocode 0, CEN EN-1990:2002, par. 4.1.2(7))

> *A nominal value, which may be specified in cases where a statistical distribution is not known.*

In Eurocode 0 (CEN EN-1990:2002, par. 1.5) a *nominal value* is defined as a

> *value fixed on non-statistical bases, for instance on acquired experience or on physical conditions*

For seismic action, which is a special accidental action unless otherwise stated, Eurocode 0 requires that (Eurocode 0, CEN EN-1990:2002, par. 4.1.2(9)) the *design value*

> *design value $A_{Ed}$ should be assessed by the characteristic value $A_{Ek}$ or specified for individual projects.*

Gulvanessian et al. (2002) underline that:

> *Note that some variable actions may not have a periodical character similar to climatic or traffic actions and the above concepts of reference period and return period may not be suitable. In this case the characteristic value of a variable action may be determined in a different way, taking into account its actual nature.*

As it has been much debated elsewhere (e.g. PAGEOPH Topical Volume 168, "Advanced Seismic Hazard Assessment" (2011) and references therein; Wyss et al, 2012; Bela, 2014), the use of historically tuned statistical indices for seismicity, based on the erroneous concept of "return period" and Poisson's statistical distributions, is rootless and unsafe, and, in particular, NDSHA does not use such assumptions. However, Eurocode 0 allows that for individual projects, which are one of the intrinsic abilities of NDSHA, design values may be otherwise specified.

It must be underlined that for variable actions, the code allows for the use of *nominal* values as *characteristic* values, that is, values that are notionally agreed because considered safe. In what follows it will be shown that the use of a maximum reference magnitude increased by a constant term $\gamma_{EM}\sigma_M$ is deeply rooted within the standard Eurocode procedures, and it is not an erratic tentative procedure.

Using the currently accepted Eurocode paradigm, in order to get the *design* value for variable actions, a further safety factor, named $\gamma_q$, is applied to the *characteristic* value of an action $Q_k$, so that the design value for the action is $Q_d=\gamma_q Q_k$. Usually, for the so-called ultimate limit states, and for typical actions like those of wind or snow, $\gamma_q=1.5$. As it has been seen, $Q_k$ may also be a nominal value.

The mechanism of fault slip giving rise to seismic waves that, after propagation from source to site, will load the structures as *accidental actions*, is governed by several parameters, among which it is mainly important, at the fault, the mechanical moment exchanged between the two sides of the slipping fault, $M_0$. In considering seismicity, at a given fault, the idea to get the *characteristic* mechanical-moment value by a statistical distribution must be abandoned due to the lack of data. The missing characteristic mechanical-moment $M_k$ acting at the fault is replaced with the *estimated* or *maximum* mechanical-moment $M_0$ acting at a given fault, that is a nominal value using the Eurocode nomenclature. The mechanical moment $M_0$ at the fault is chosen according to what follows, in NDSHA:

a) The mechanical moment $M_0$ considered is a reasonable lower estimate of the worst that might physically happen when only seismogenic nodes are used, since, by definition, they accommodate earthquakes with a magnitude above a fixed threshold (Gorshkov et al., 2002; 2004; Peresan et al., 2015) (this is NDSHA suggested procedure when no historical catalogues are available or they are not considered sufficiently complete).
b) The mechanical moment $M_0$ considered is the maximum in the parametric catalogue if only the catalogue and seismogenic zones are used (this is the original, chronologically first NDSHA).
c) The mechanical moment considered is the maximum between parametric catalogue and seismogenic nodes, all within seismogenic zones, if are all used at the same time (this is NDSHA as suggested in regions where catalogues are available).

The mechanical moment applied at the fault is one of the main input of NDSHA, at a given source. Then, by also controlling other parameters, NDSHA is able to fully simulate waves induced in the earth-mechanical system along the path from source to the bedrock at the site, and with site specific analyses, the path from the bedrock at the site to the surface at the site.

In other words, the mechanical moment $M_0$ acts, within the framework of NDSHA, as a variable action whose (usually maximum) value is estimated in order to compute some mechanical effects caused by it. The analogy with what is done by structural engineers with their systems is strict, in theoretical sense.

The moment magnitude $M_w$ is related to the mechanical moment acting at the fault $M_0$ by the well-known Hanks-Kanamori formula (Hanks and Kanamori, 1979), where the mechanical moment is measured in Nm:

$$M_W = \frac{2}{3}\left[\text{Log}(M_0)-6\right]$$

If the same rule used by Eurocode 0 for variable actions is applied to the mechanical-moment acting at the fault $M_0$, which acts as an input action $Q_k$, we get

$$M_{W,design} = \frac{2}{3}\left[\text{Log}(\gamma_q M_0) - 6\right]$$

Which leads to

$$M_{W,design} = \frac{2}{3}\left[\text{Log}(\gamma_q M_0) - 6\right] = M_W + \frac{2}{3}\text{Log}(\gamma_q)$$

So in order to factor the mechanical-moment acting at the fault, which acts as a mechanical generalized force applied to the system, the magnitude related to it should be increased by a *fixed* increment, namely

$$\Delta M = \frac{2}{3}\text{Log}(\gamma_q)$$

If we set this increment equal to $\sigma_M$ times a new safety factor $\gamma_{EM}$, we get

$$\gamma_{EM}\sigma_M = \frac{2}{3}\text{Log}(\gamma_q)$$

Assuming $\sigma_M=0.2$ and $\gamma_{EM}$ in the range 1.5–2.5 (e.g. Rugarli et al., 2018) we get that $\gamma_q$ can be defined in the integer range $\gamma_q=$3–6. The high values, when compared to the current 1.5 used for more friendly actions like wind or snow, or passing vehicles, are the counterpart of the much higher uncertainness related to earthquakes. In fact $\gamma_q=$3–6 is well consistent with the variation that may affect $M_0$, as determined, for the same event, by different agencies and methods (e.g. Panza and Saraò, 2000; Saraò et al., 2001; Guidarelli and Panza, 2006).

So when the estimated magnitude at a fault is increased by a constant value, this is equivalent to factoring the mechanical moment acting at the fault, exactly as a structural engineer following the format of partial safety factor method would do.

**3 - Validation of the safety factor**

NDSHA's aim is to supply an *envelope value*, in other words *a value that should not be exceeded*, therefore it is immediately falsifiable: if an earthquake occurs with a magnitude $M_{eq}$, *larger* than that indicated by NDSHA's $M_{design}$, then $\Delta M=M_{eq}-M_{max}>\gamma_{EM}\sigma_M$ and $\gamma_{EM}$ should be increased. Given the way $M_{design}$ is defined, however, this is expected to be a rare condition.

$\gamma_{EM}$ should similarly be increased, should recorded peak ground motion values (e.g. PGA) on the bedrock at the occurrence of an earthquake $M_{eq}$ *after* the compilation of NDSHA maps, exceed within error limits those values given in these same maps. By way of improving usefulness and applicability of future strong ground motion recordings, this would suggest to possibly install additional stations over stiff soils, so as to avoid the *local amplifications* due to site effects. Today the majority of the strong ground motion stations of for instance the Italian net, are sited over soft soils.

The selection of the multiplier $\gamma_{EM}$ to be applied to the standard deviation cannot be proved by equations, and it would be misleading to try to do so. Therefore the choice of its value is partly heuristic, or rule-of-thumb. Nonetheless, should this heuristic be falsified by natural experiments, this multiplier can be gradually reset to the minimum safe value. This is what has already been done with all the safety factors used in engineering: (i) the $\gamma=1.5$ safety factor for material limit stresses was used well before the availability of reliable statistical measures; and (ii) the semi-probabilistic methods used in structural engineering are *de facto* tuned to confirm these already validated-by-experience values (Rugarli et al., 2018).

**4 - Tuning of $\gamma_{EM}$ in Italy**

Italy is the country with the longest parametric earthquake catalogue based upon both historical and instrumental data. As it is well known the loglinear Gutenberg-Richter relation (GR) represents a law only at global scale (Båth, 1973; Kosobokov and Mazhkenov, 1994; Molchan et al., 1997).

Considering the loglinear GR relation as a law, the used Italian earthquake catalogue can be considered sufficiently *complete* (e.g. Vorobieva and Panza, 1993) at national scale, starting from year 1000, for events with magnitude $M$ such that $M_{tr}<M<M_{up}$ that is, for Italy, $5.0<M<7.5$.

The magnitude $M_{tr}=5$ is the lower magnitude threshold used for NDSHA computations (Panza et al., 2001). The upper magnitude $M_{up}$ is related to the specific of Italian territory.

Completeness according to GR means that:

1. in the catalogue the number of occurred but missed earthquakes having $M>M_{tr}$, that is the *completeness threshold* (in our case $M_{tr}=5.0$), is minor;
2. the information content of the catalogue cannot exclude the occurrence, in Italy, of a future event with $M>M_{up}=7.5$.

Naturally, completeness does not obviously imply ubiquitous representativeness of the real earthquake hazard and this imposes special care in the definition of MCE. *"As far as the laws of mathematics refer to reality, they are not certain; and as far as they are certain, they do not refer to reality" (Albert Einstein, Geometry and Experience - an expanded form of an Address to the Prussian Academy of Sciences in Berlin on January 27th, 1921).*

Via the GR law, the extension of the region to be considered and the interval of magnitudes are related, as the sources in all the magnitude range must be so small, when compared to the region extent, to be considered as points: GR law is only valid when considering a region sufficiently extended.

Having a catalogue that might be considered *complete* in a given area and for a given magnitude interval, implies having some root in assuming that the still not experienced faults are a small number, otherwise major, not minor differences between the expected and the historically recorded number of events could be estimated, should GR law be considered valid.

If this is true, then the catalogue is a useful tool to tune the safety factor $\gamma_{EM}$ in a procedure where only *seismogenic nodes* are used, *and not the catalogue itself or the related seismogenic zones*.

In this way the catalogue is used as a huge set of experiments, to be tested against some possibly assumed safety factors $\gamma_{EM}$. Full independency between the map considered summing a given constant $\gamma_{EM}\sigma_M$ to the magnitude of seismogenic nodes, and the results of the catalogue, is preserved. The two sets are totally independent.

**4.1 - Hazard map based on seismogenic zones and parametric earthquake catalogue**

This map is the map used as target value, i.e. the map has been considered equivalent to a set of experimental results, during 1000 years. This is a unique-worlwide set of data. However, the *experimental* macroseismic intensities, and then the magnitudes and derived $M_0$, are necessarily affected by high uncertainness, that should be considered when using them.

The magnitude smoothing procedure applied by NDSHA for regional scale computations (Panza et al., 2001) is a first step aimed at a conservative earthquake hazard estimate. Figure 1 shows the earthquake sources defined within the ZS9 seismic zones (Meletti et al., 2008), before (left) and after (right) the smoothing window of three cells has been applied to the magnitude values reported in the CPTI04 parametric catalogue (CPTI04 Working Group, 2004), discretized into 0.2°x0.2° cells. No magnitude increment $\gamma_{EM}\sigma_M$ has been applied to this map.

The NDSHA map of Design Ground Acceleration (DGA) based on the smoothed magnitude distribution is shown in Figure 2. It corresponds to Model 3 of Panza et al. (2012), but for sake of simplicity, the Size Scaled Point Source model (SSPS) has been used here instead of the Size and Time Scaled Point Source one (STSPS) (Panza et al., 2012).

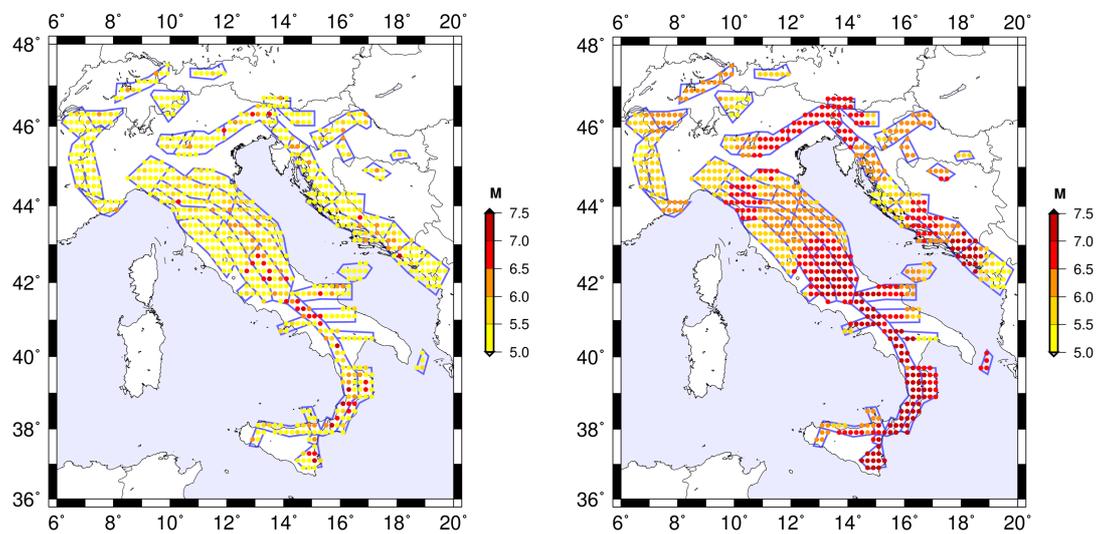

Figure 1. Earthquake sources used for the computation of the synthetic seismograms, which are at the base of the NDSHA maps at regional scale ($\gamma_{EM}$=0). For our study area, the sources are located within the seismogenic zones defined by ZS9 (Meletti et al., 2008), with additions by A.A. V.V. (2001). Left: distribution obtained from the unperturbed magnitudes available in the CPTI04 (CPTI04 Working Group, 2004) earthquake catalogue, discretized into 0.2°x0.2° cells. Right: distribution obtained after the smoothing procedure has been applied.

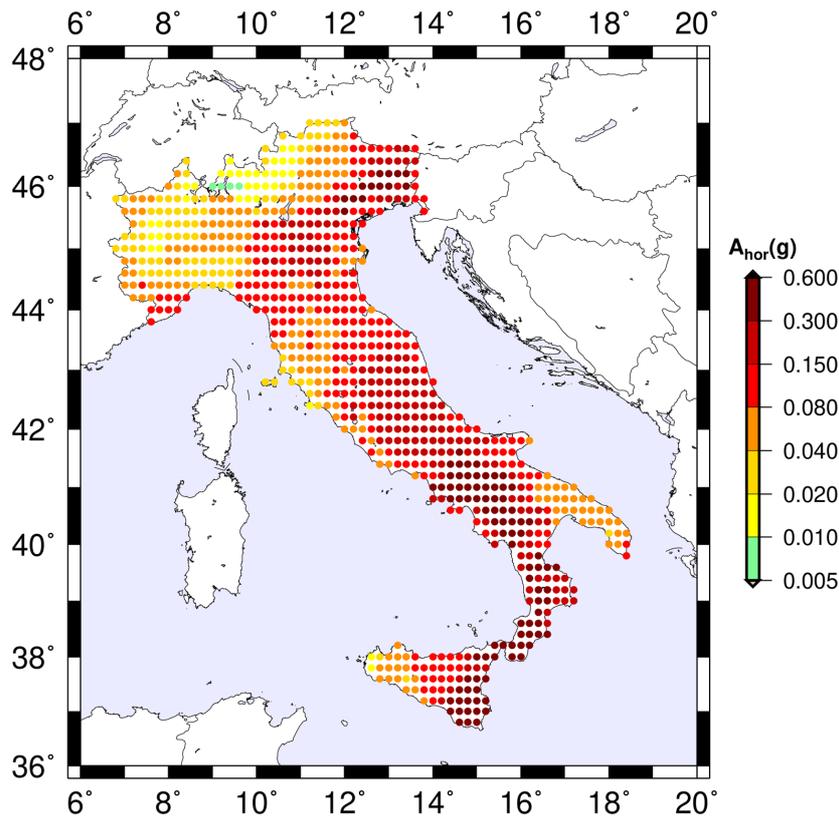

Figure 2. Horizontal DGA map computed using the earthquake sources obtained after smoothing the unperturbed discretized magnitude of catalogue CPTI04 (Figure 1, right). This may be considered as the map got by a 1000 years lasting set of experiments.

**4.2 - Hazard map based on seismogenic nodes**

This map (see Figure 3) is the result of using NDSHA considering only seismogenic nodes, and applying a safety factor $\gamma_{EM}$ as explained in section 2.

The seismogenic nodes are defined by the morphostructural zonation (MSZ) based on pattern recognition techniques (Alekseevskaya et al.,1977, Peresan et al., 2011). The method for the pattern recognition of earthquake-prone areas is based on the assumption that strong events nucleate at the morphostructural nodes (Gelfand et al., 1972), specific structures that are formed at the intersections of lineaments.

Talwani (1988, 1999) found that large intraplate earthquakes are related to intersections of lineaments and proposed a model demonstrating that intersecting faults provide a location for stress accumulation. Hudnut et al. (1989) and Girdler and McConnell (1994) evidenced the relationship between earthquakes and intersections for plate boundaries and rift structures, respectively.

According to King (1986), fault intersection zones provide locations for initiation and cessation of ruptures. The non-randomness of earthquake nucleation at the nodes is proved statistically by a specifically designed method (Gvishiani and Soloviev, 1981).

Recent estimations of the validity of the worldwide recognition results of earthquake-prone areas are given by Gorshkov and Novikova (2018) who report a global score of about 86%. Such a value confirms earlier investigations about the percentage of post-publication earthquakes falling in the recognized seismogenic nodes (Soloviev et al., 2014; Peresan et al., 2015).

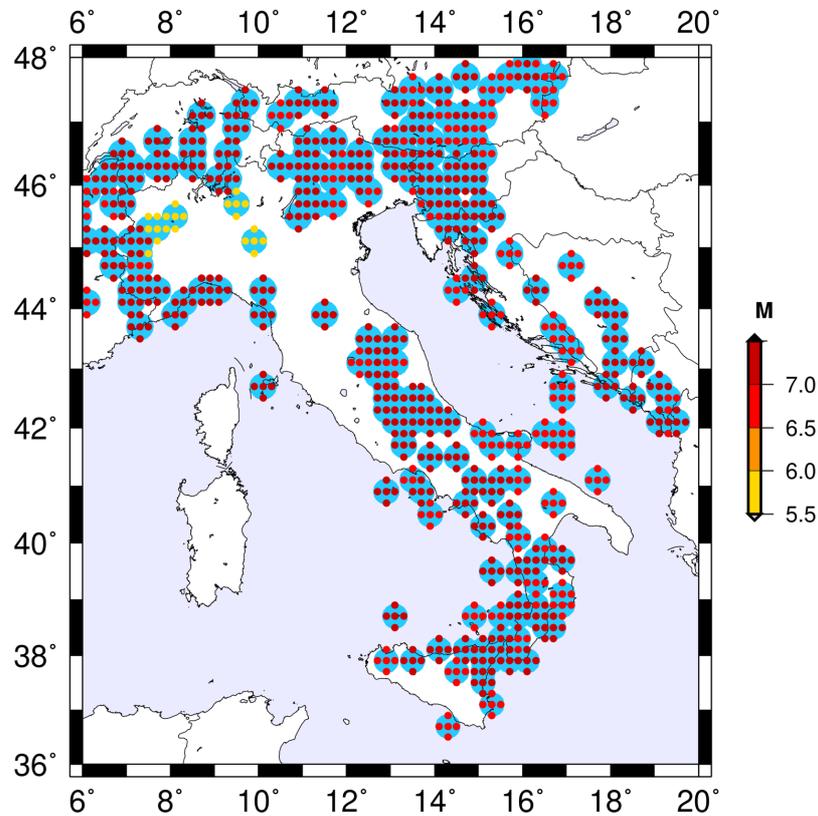

Figure 3. Earthquake sources used for the computation of synthetic seismograms, defined within the seismogenic nodes, with magnitude $M_{sz}$ increased by 0.5, i.e. $\Delta M = \gamma_{EM}\sigma_M = 0.5$. The info coming from the historical catalogue of seismicity is here deliberately *neglected*, to evaluate the hazard forecasting capability supplied by the seismogenic nodes, identified by morphostructural zonation (MSZ) and pattern recognition.

The sufficient validity of the methodology for identifying areas capable of strong earthquakes is proven and, at the same time, the idea about nucleating strong earthquakes at the nodes is confirmed, as recently empirically observed by Walters et al. (2018) in the 2016 Central Italy seismic sequence. Such validation is very important since the model of seismogenic nodes is mathematically rigorous, based upon objective, but not error-free morphostructural data.

A morphostructural map has been compiled for Italy by Gorshkov at al. (2002; 2004). Under the assumption that future strong events will occur at the nodes, they evaluated the seismic potential of each node by means of the pattern recognition technique for two magnitude thresholds: $M_{sz} \geq 6.0$ and $M_{sz} \geq 6.5$. The nodes prone to earthquakes with $M_{sz} \geq 6.0$ are identified with the pattern recognition algorithm "CORA-3" (Gelfand et al., 1972). The nodes with larger earthquakes potential are identified by the criteria of high seismicity derived by Kossobokov (1983) from pattern recognition in the Pamirs-Tien Shan region.

For recognition purposes, the nodes have been defined as circles of radius R=25 km surrounding each point of intersection of lineaments (Gorshkov at al., 2002). Such node dimension is comparable with the size of the earthquake source for the magnitude range considered in this work (Wells and Coppersmith, 1994).

In Figure 3 the earthquake sources are shown that, *ignoring both the ZS9 seismic zones and the CPTI04 catalogue*, are obtained considering *only the seismogenic nodes*.

The reference magnitude used for the earthquake at a seismogenic node, is a lower bound of the earthquake magnitude threshold, $M_{sz}$, identified for that node by the morphostructural analysis (Gelfand et al., 1972).

In Figure 3 the shown magnitude is $M_{sz}$ incremented by $\Delta M$=0.5, i.e. $2\sigma_M$ (e.g Båth, 1973, p. 111). To the published information (Gorshkov et al., 2002; 2004) a few nodes with $M_{sz}$=5.0, identified in the western Po plain using "CORA-3", have been considered (Peresan et al., 2015), again after raising by 0.5 their magnitude. The corresponding map of DGA is given in Figure 4.

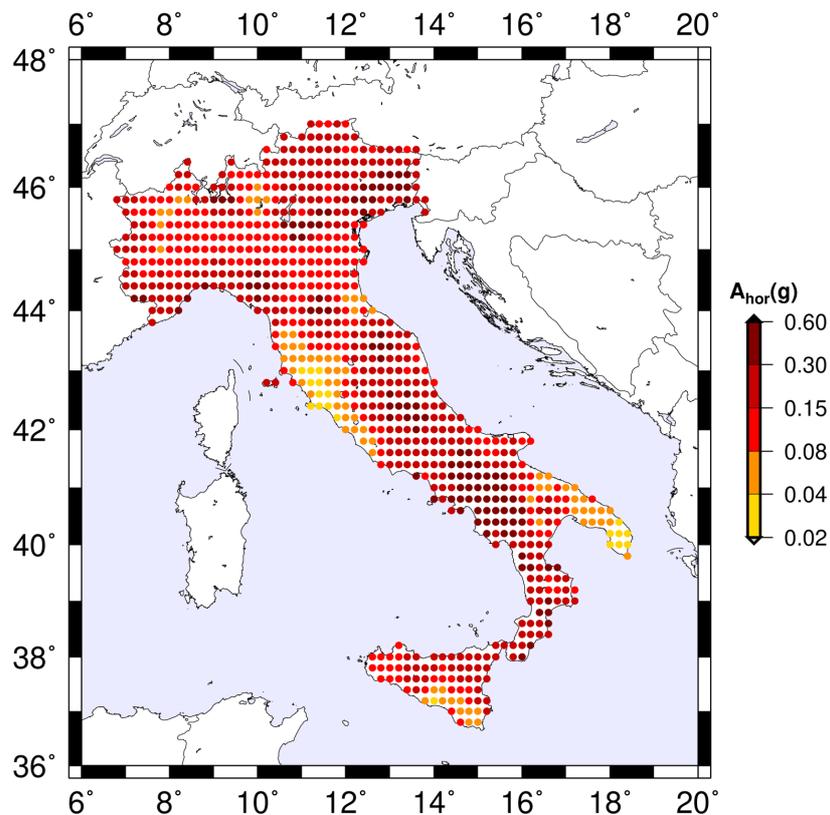

Figure 4. Horizontal DGA map computed using the earthquake sources defined within the seismogenic nodes (see Figure 3), with magnitude $M_{sz}$ incremented by 0.5, i.e. $\Delta M=\gamma_{EM}\sigma_M$=0.5.

## 5 - Discussion

In Figure 5 the ratio between the values of DGA given in Figure 2 (sources defined processing the magnitudes found in the historical catalogue of seismicity CPTI04) and those in Figure 4 (sources defined within the seismogenic nodes and using $\gamma_{EM}=2.0$) is shown.

It must be kept in mind that in the figure real differences are those greater than 2; a factor of 2 is comparable with the resolving power of the available experimental data about *M*, essentially all based on Macroseismic Intensity ($I_{MCS}$). In fact, any intensity scale is *discrete* and therefore it has *unit* incremental steps; intermediate values are not defined. Typical discrete ranges of hazard values (units of g) are in geometrical progression (close to 2), consistent with the real resolving power of the worldwide available experimental data (e.g. Cancani, 1904; Lliboutry, 2000).

The empty-of-triangles regions are those where the two maps match in terms of intensity; the match can still be considered satisfactory in correspondence of the light blue triangles ($\Delta I=1$) when considering the error threshold intrinsically related to macroseismic intensity $I_{MCS}$.

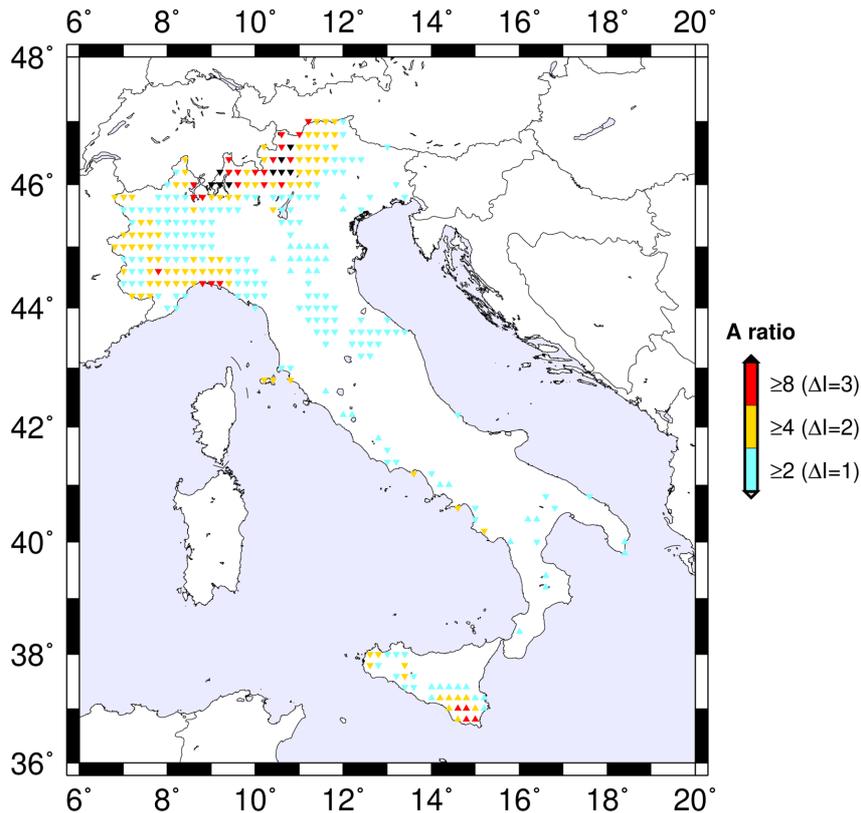

Figure 5. Map of the ratios between the Design Ground Acceleration (DGA) values given in Figure 2 and Figure 4. The upward triangles indicate larger values obtained considering the CPTI04 earthquake catalogue, while the downward triangles indicate larger values obtained considering the sources associated with the seismogenic nodes, whose magnitude has been increased by 0.5, i.e. $\Delta M=\gamma_{EM}\sigma_M=0.5$.
11

From Figure 5 it is evident that for $\Delta M=\gamma_{EM}\sigma_M=0.5$, or equivalently $\gamma_{EM}=2.0$, the hazard at the bedrock, defined only considering seismogenic nodes, *safely and reliably envelopes peak values observed in the last millennium almost everywhere in Italy*. We believe this is a remarkable result, considering that the two maps are totally independent one another. Almost everywhere the two maps match, i.e. the difference in DGA ratio is lower than 2. There are two exceptions:

1. a few locations in southeastern Sicily;
2. Central Alps

that will be now discussed.

The space distribution of a few locations in southeastern Sicily nicely mirrors the portion of the Africa plate occupied by the Hyblaean Mountains, south of the Apennines arcuate thrusts front (e.g. Carminati and Doglioni, 2012; Doglioni and Panza, 2015); there a larger value of $\Delta M=\gamma_{EM}\sigma_M=0.5$ seems to be necessary. This increment, while keeping $\gamma_{EM}=2.0$, can be consistently obtained assuming for $\sigma_M$ the upper bound of the range estimated by Båth (1973) for instrumental magnitudes, and nowadays routinely confirmed by global estimates of $M$ supplied for the same earthquake by different Agencies. However, a test done with $\Delta M=\gamma_{EM}\sigma_M=0.6$ still underestimates the ground shaking obtained using the historical catalogue in southeastern Sicily.

The large differences that can be read in the map for southeast Sicily, and that could be explained by the possibly natural discrepancy between the distribution of earthquake sources within the seismogenic zones, on one side, and the seismogenic nodes, on the other, can hardly be attributed to the local mechanical properties.

It must be considered the high uncertainness of the historical catalogue when evaluating macroseismic intensity of specific events. When *just one event* is determinant, this is particularly true, because a single possible error might directly impact the mapped final result (of Figure 2).

The use of a $\sigma_M \gg 0.3$ seems to be appropriate in this particular case. Indeed for the 1693 M=7.4 event, that seems to control the hazard in the zone, $\sigma_M=0.7$ can be assumed, a typical value for magnitudes derived from $I_{MCS}$ according to D'Amico et al. (1999), also on account of the proximity of the epicentral area to the sea. This means that the uncertainness embedded in $\sigma_M$ may be increased to 0.7 instead of using the normal 0.2−0.3, for the particular nodes referring to the region where the 1693 event was recorded, still being within the error limit acceptable.

Therefore it is consistent to add to $M_{sz}$ a value of $\Delta M=\gamma_{EM}\sigma_M=1.4$ and thus consider a magnitude value 7.4 in the computation of DGA using the earthquake sources defined within the two southeasternmost seismogenic nodes in the area (nodes 142 and 145 in Gorshkov et al., 2002). In such a way, i.e. within experimental errors, the underestimates in southeastern Sicily, shown in Figure 5, fade away as shown in Figure 6.

The minimum error affecting the *experimental* $I_{MCS}$ is, by definition, one degree, as well confirmed empirically for 55 damaging earthquakes occurred in Italy since 461, as shown by the detailed analysis made by Kronrod et al. (2002). Therefore it can be concluded that, within errors, the hazard assessed considering only the seismogenic nodes envelopes (exceeds or equals) the one assessed considering CPTI04 catalogue and ZS9 seismic zonation, and almost everywhere does indeed match it.

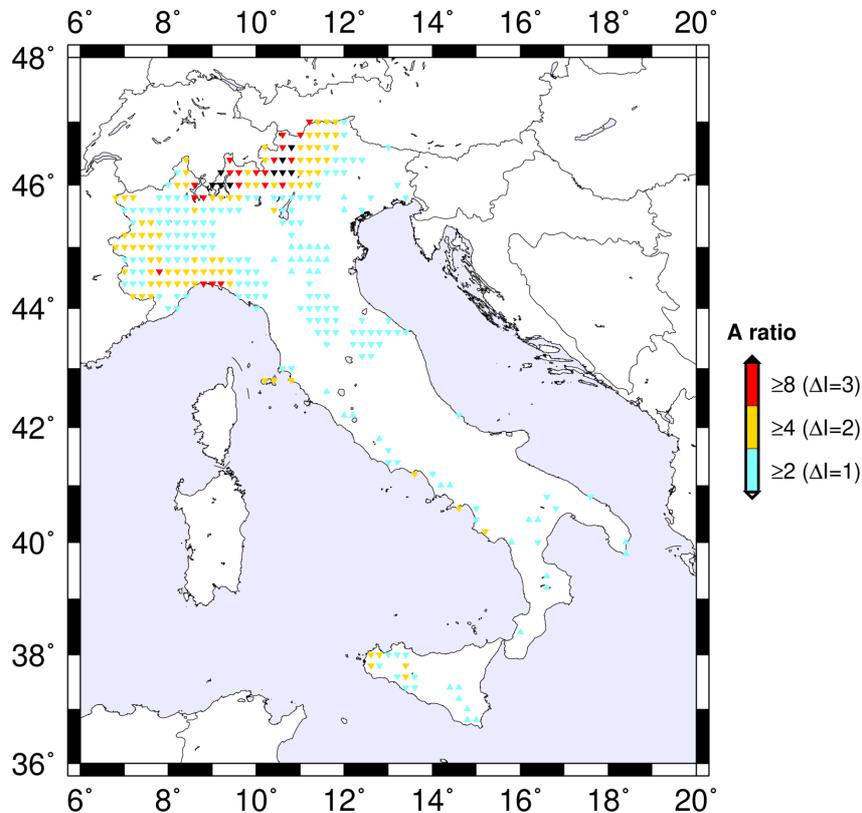

Figure 6. Same as Figure 5 but with $\Delta M = \gamma_{EM}\sigma_M = 1.4$ in southeast Sicily.

On the other side, $\Delta M = \gamma_{EM}\sigma_M = 0.5$ leads to some diffused overestimation in Northern Italy, with a relevant peak in Central Alps, that remains outstanding even using of $\Delta M = \gamma_{EM}\sigma_M = 0.4$. Values of $\Delta M = \gamma_{EM}\sigma_M > 0.6$, required to eliminate the hazard underestimation in southeastern Sicily, would of course imply much larger overestimations elsewhere, with the largest discrepancies obviously in the Central and Western Alps.

Thus, it may be argued that part of the observed misfit can be naturally explained by a different level of error affecting empirical $M$ determinations. The only exception is limited to the Central Alps, an area where the >1000 years long catalogue CPTI04, even though complete for $M$>5 at national scale (Vorobieva and Panza, 1993), may not fully represent the local seismicity, given that there the occurrence rate (i.e. the normalized count of events in the observation time) of large earthquakes ($I_{MCS}$>VII) seems to be smaller than 1 event per 1000 years (Magrin et al., 2017; Michetti et al., 2012; Houlié et al., 2018).

This may imply an event of significant magnitude that is not present in today's historical catalogue for this area or could be explained by the occurrence of dislocation creep within a lithosphere with fairly anomalous mechanical and rheological properties (Panza and Raykova, 2008; Malusà et al., 2018).

It must be underlined that the two regions discussed up to this point are very limited when compared to the extension of Italy.

## 6 - Conclusions

It has been shown that incrementing by a constant variation $\Delta M$ the magnitude considered at a fault, in order to get safe envelope of seismic actions, is totally equivalent to applying a partial safety factor $\gamma_q$ to the mechanical moment at the fault $M_0$, as normally asked by Eurocodes and other international standards for typical structural actions ($\Delta M = \frac{2}{3}\text{Log}(\gamma_q)$).

In turn, this is equivalent to applying a $\gamma_{EM}$ safety factor to the typical standard deviation in the evaluation of earthquake magnitude, $\sigma_M$, so as to get the desired magnitude increment $\Delta M = \gamma_{EM}\sigma_M$, constant for each seismic source.

By using as typical value $\gamma_{EM}\sigma_M=0.5$ ($\sigma_M=0.25$, central value of the experimental range 0.2–0.3, and $\gamma_{EM}=2$) and NDSHA using only seismogenic nodes, it has been shown that the unique 1,000-year long Italian earthquake catalogue, acting as experimental set, is, within errors, almost everywhere matched or enveloped.

Minor variations are related to possible error in evaluating $I_{MCS}$ of historical events, or in *missing* seismic events in a very restricted Italian area, with very peculiar mechanical and rheological properties.

## 7 - References

A.A. V.V., 2000. Seismic hazard of the Circum-Pannonian region, (G. F. Panza, M. Radulian and C.-Y. Trifu editors), PAGEOPH Topical Volumes, 157, Birkhouser, Basel.

Alekseevskaya, M.A., Gabrielov, A.M., Gvishiani, A.D., Gelfand, I.M. and Ranzman, E. Y. (1977). Formal morphostructural zoning of mountain territories. J. Geophys., 43, 227–233.

Bela, J. (2014). Too generous to a fault? Is reliable earthquake safety a lost art? Errors in expected human losses due to incorrect seismic hazard estimates, Earth's future, 2, 569–578, DOI: 10.1002/2013EF000225.

Båth, M. (1973). Introduction to Seismology. Birkhäuser Verlag, Basel.

Cancani, A. (1904). Sur l'emploi d'une double échelle sismique des intensités, empirique et absolue. Gerlands Beitr. Geophys., 2, 281–283.

Carminati, E. and Doglioni, C. (2012). Alps vs Apennines: the paradigm of a tectonically asymmetric Earth. Earth Science Reviews, 112, 67–96. http://dx.doi.org/10.1016/ j.earscirev.2012.02.004.

CEN EN 1990:2002 Eurocode - Basis of Structural Design, April 2002.

CPTI Working Group (2004). Catalogo Parametrico dei Terremoti Italiani, versione 2004 (CPTI04). Bologna, Italy.

D'Amico, V., D. Albarello and Mantovani, E. (1999). A distribution-free analysis of magnitude-intensity relationships: an application to the Mediterranean region. Physics and Chemistry of the Earth, Part A: Solid Earth and Geodesy 24, 517–521.

Doglioni, C. and Panza, G.F. (2015). Polarized plate tectonics. Advances in Geophysics, 56, 1–167, Elsevier, ISSN: 0065-2687, Doi: 10.1016/bs.agph.2014.12.001.

Gelfand, I.M., Guberman, S.I., Izvekova, M.L., Keilis-Borok, V.I. and Ranzman, E.J. (1972). Criteria of high seismicity, determined by pattern recognition. Tectonophysics, 13, 415–422.

Girdler, R.W. and McConnell, D.A. (1994). The 1990 to 1991 Sudan earthquake sequence and the extent of the East African Rift System. Science, 264, 67–70.

Gorshkov, A. and Novikova, O. (2018). Estimating the validity of the recognition results of earthquake-prone areas using the ArcMap. Acta Geophys. https://doi.org/10.1007/s11600-018-0177-3, Online ISSN 1895-7455.

Gorshkov, A.I., Panza, G.F., Soloviev, A.A. and Aoudia, A. (2002). Morphostructural zonation and preliminary recognition of seismogenic nodes around the Adria margin in peninsular Italy and Sicily. J. Seismol. Earthq. Eng., 4, 1–24.

Gorshkov, A.I., Panza, G.F., Soloviev, A.A. and Aoudia, A. (2004), Identification of seismogenic nodes in the Alps and Dinarides. Boll. Soc. Geol. It., 123, 3–18.

Guidarelli, M. and Panza, G.F. (2006). INPAR, CMT and RCMT seismic moment solutions compared for the strongest damaging events ($M³4.8$) occurred in the Italian region in the last decade. Rend. Accad. Naz. delle Scienze detta dei XL Mem. di Scienze Fisiche e Naturali, 30, pp. 81–98.

Gulvanessian H., Calgaro J-A. and Holicky, M. (2002). Eurocode: Basis of Structural Design, Designers' Guide to Eurocodes, Thomas Telford.

Gvishiani, A.D. and Soloviev, A.A. (1981). Association of the epicenters of strongearthquakes with the intersections of morphostructural lineaments in South America. In V.I. Keilis-Borok and A.L. Levshin (ed.) Interpretation of seismic data: methods and algorithms. Comput. Seismol. 13, Allerton, New York, 42–46.

Hanks, T.C. and Kanamori, H., (1979). A Moment Magnitude Scale. J.G.R., 84, B5, May 10, 1979.

Hassan, H.M., Romanelli, F., Panza G.F., El Gabry, M.N. and Magrin, A. (2017). Update and sensitivity analysis of the neo-deterministic seismic hazard assessment for Egypt. Eng. Geol., 218, 77–89, http://dx.doi.org/10.1016/j.enggeo.2017.01.006.

Houlié, N., Woessner, J., Giardini, D. and Rothacher, M. (2018). Lithosphere strain rate and stress field orientations near the Alpine arc in Switzerland. Nature Scientific Reports, 8:2018, DOI:10.1038/s41598-018-20253-z.

Hudnut, K.W., Seeber, L. and Pacheco, J. (1989). Cross-fault triggering in the November 1987 Superstition Hills earthquake sequence, Southern California. Geophys. Res. Lett., 16, 199–202.

King, G. (1986). Speculations on the geometry of the initiation a termination processes of earthquake rupture and its relation to morphology and geological structure. Pure and Applied Geophysics, 124, 567–583.

Kosobokov, V.G. and Mazhkenov, S.A. (1994). On similarity in the spatial distribution of seismicity. In D.K. Chowdhury (ed.), Computational Seismology and Geodynamics / Am. Geophys. Un., 1, Washington, D.C., 6–15.

Kossobokov, V.G. (1983). Recognition of the sites of strong earthquakes by Hamming's method in East Central Asia and Anatolia, In V.I. Keilis-Borok and A.L. Levshin (eds), Computational Seismology, 14, Allerton Press, Inc., New-York, 78–82.

Kronrod, T.L., Molchan, G.M., Podgaetskaya, V.M. and Panza, G.F. (2002). Formalized representation of isoseismal uncertainty for Italian earthquakes. Bollettino di Geofisica Teorica ed Applicata, 41, 243–313.

Lliboutry, L. (2000). Quantitative geophysics and geology, Springer-Verlag, London, UK, ISBN 978-1-85233-115-3.

Magrin, A., Peresan, A., Kronrod, T., Vaccari, F. and Panza, G.F. (2017). Neo-deterministic seismic hazard assessment and earthquake occurrence rate. Eng. Geol., 229, 95–109.

Malusà, M.G., Frezzotti, M.L., Ferrando, S., Brandmayr, E., Romanelli, F. and Panza, G.F. (2018). Active carbon sequestration in the Alpine mantle wedge and implications for long-term climate trends. Nature Scientific Reports, 8:4740, DOI:10.1038/s41598-018-22877-7.

Meletti, C., Galadini, F., Valensise, G., Stucchi, M., Basili, R., Barba, S., Vannucci, G. and Boschi, E. (2008). A seismic source zone model for the seismic hazard assessment of the Italian territory. Tectonophysics, 450, 85–108.

Michetti, A., Giardina, F., Livio, F., Mueller, K., Serva, L., Sileo, G., Vittori, E., Devoti, R., Riguzzi, F., Carcano, C., Rogledi, S., Bonadeo, L., Brunamonte, F. and Fioraso, G. (2012). Active compressional tectonics, Quaternary capable faults, and the seismic landscape of the Po Plain (northern Italy). Ann. Geophys., 55, 5, doi: 10.4401/ag-5462.

Molchan, G., Kronrod, T. and Panza, G.F. (1997). Multi-scale seismicity model for seismic risk. Bull. Seismol. Soc. Am., 87, 1220–1229.

PAGEOPH Topical Volume 168 (2011) "Advanced seismic hazard assessment, Vol. 1 and Vol. 2", Editors: Panza G.F., Irikura K., Kouteva-Guentcheva M., Peresan A., Wang Z. and Saragoni R., Pure Appl. Geophys., Birkhäuser, Basel, Switzerland, Vol. 1, ISBN 978-3-0348-0039-6, http://www.springer.com/it/book/9783034800396, Vol. 2, ISBN 978-3-0348-0091-4, http://www.springer.com/it/book/9783034800914.

Panza, G.F., La Mura, C., Peresan, A., Romanelli, F. and Vaccari, F. (2012). Seismic hazard scenarios as preventive tools for a disaster resilient society. Advances in Geophysics, 53, 93–165.

Panza, G.F. and Raykova, R.B. (2008). Structure and rheology of lithosphere in Italy and surrounding. Terra Nova, 20, 194–199.

Panza, G.F., Romanelli, F. and Vaccari, F. (2001). Seismic wave propagation in laterally heterogeneous anelastic media: theory and applications to seismic zonation. Advances in Geophysics, 43,1–95.

Panza, G.F. and Saraò, A. (2000). Monitoring volcanic and geothermal areas by full seismic moment tensor inversion: are non-double couple components always artifacts modeling? Geophys. J. Int., 143, 353–364.

Parvez, I.A., Magrin, A., Vaccari, F, Ashish, Mir, R.R., Peresan, A. and Panza, G.F. (2017). Neo-deterministic seismic hazard scenarios for India - a preventive tool for disaster mitigation. J. Seismol., doi 10.1007/s10950-017-9682-0.

Peresan, A., Gorshkov, A., Soloviev, A. and Panza, G.F. (2015). The contribution of pattern recognition of seismic and morphostructural data to seismic hazard assessment. Bollettino di Geofisica Teorica ed Applicata, 56, 295–328; DOI 10.4430/bgta0141.

Peresan, A., Zuccolo, E., Vaccari, F., Gorshkov, A. and Panza, G.F. (2011). Neo-deterministic seismic hazard and pattern recognition techniques: Time-Dependent Scenarios for North-Eastern Italy. Pure Appl. Geophys., 168, 583–607; doi: 10.1007/s00024-010-0166-1.

Rugarli, P., Amadio, C., Peresan, A., Fasan, M., Vaccari, F., Magrin, A., Romanelli, F. and Panza, G.F., (2018). Neo-deterministic scenario-earthquake accelerograms and spectra: a NDSHA approach to seismic analysis, in: Jia, J. and Paik, J.K (Eds) Structural Engineering in Vibrations, Dynamics and Impacts, CRC press, Taylor & Francis Group, Abingdon, UK, in press.

Saraò, A., Panza, G.F., Privitera, E. and Cocina, O. (2001). Non-double couple mechanisms in the seismicity preceding 1991–1993 Etna volcano eruption. Geophys. J. Int., 145, 319–335.

Soloviev, A.A., Gvishiani, A.D., Gorshkov, A.I., Dobrovolsky, M.N. and Novikova, O.V. (2014). Recognition of earthquake-prone areas: methodology and analysis of the results. Izvestiya, Physics of the Solid Earth, 50, 151–168, DOI: 10.1134/S10693513140201.

Talwani, P. (1988). The intersection model for intraplate earthquakes. Seismol. Res. Lett., 59, 305–310.

Talwani, P. (1999). Fault geometry and earthquakes in continental interiors. Tectonophysics, 305, 371–379.

Vorobieva, I.A. and Panza, G.F. (1993). Prediction of the occurrence of related strong earthquakes in Italy. Pure Appl. Geophys., 141, 25–41.

Walters, R.J., Gregory, L.C., Wedmore, L.N.J., Craig, T.J., McCaffrey, K., Wilkinson, M., Chen, J., Li, Z., Elliott, J.R., Goodall, H., Iezzi, F., Livio, F., Michetti, A.M., Roberts, G. and Vittori, E. (2018). Dual control of fault intersections on stop-start rupture in the 2016 Central Italy seismic sequence. Earth and Planetary Science Letters, 500 1–14, https://doi.org/10.1016/j.epsl.2018.07.043.

Wells, D.L. and Coppersmith, K.J. (1994). New empirical relationships among magnitude, rupture length, rupture width, and surface displacement. Bull. Seism. Soc. Am., 84, 974–1002.

Wyss, M., Nekrasova, A. and Kossobokov, V. (2012). Errors in expected human losses due to incorrect seismic hazard estimates. Nat. Haz., 62, 927–935, http://dx.doi.org/10.1007/s11069- 012-0125-5.